# Thickness Dependent Parasitic Channel Formation at AlN/Si Interfaces

Hareesh Chandrasekar,[1,2*†] K N Bhat,[1] Muralidharan Rangarajan,[1] Srinivasan Raghavan[1,2] and Navakanta Bhat[1,3*]

[1] Centre for Nano Science and Engineering, Indian Institute of Science, Bangalore, 560012, India

[2] Materials Research Centre, Indian Institute of Science, Bangalore, 560012, India

[3] Department of Electrical Communication Engineering, Indian Institute of Science, Bangalore, 560012, India

[*] hareeshc2408@gmail.com; navakant@iisc.ac.in

[†] currently affiliated with the Centre for Device Thermography and Reliability, HH Wills Physics Laboratory, University of Bristol, Bristol, BS8 1TL, UK

The performance of GaN-on-Silicon electronic devices is severely degraded by the presence of a parasitic conduction pathway at the nitride-substrate interface which contributes to switching losses and lower breakdown voltages. The physical nature of such a parasitic channel and its properties are however, not well understood. We report on a pronounced thickness dependence of the parasitic channel formation at AlN/Si interfaces due to increased surface acceptor densities at the interface in silicon. The origin of these surface acceptors is analyzed using secondary ion mass spectroscopy measurements and traced to thermal acceptor formation due to Si-O-N complexes. Low-temperature (5K) magneto-resistance (MR) data reveals a transition from positive to negative MR with increasing AlN film thickness indicating the presence of an inversion layer of electrons which also contributes to parasitic channel formation but whose contribution is secondary at room temperatures.

1. INTRODUCTION:

The availability of large area silicon substrates at low cost, allied with the development of CMOS-compatible process flows for GaN device fabrication are rapidly enhancing the commercial viability of GaN-on-silicon electronics, making such devices cost-competitive with traditional Si-based solutions for high-power, high-frequency applications.[1-7] GaN-on-Si discrete power devices such as high electron mobility transistors (HEMTs) and diodes are expected to have a major impact in terms of switching efficiency and size (form-factor) in the 200-1200V application ranges.[8] For RF applications, GaN HEMTs on SiC deliver the best performance at a higher cost, with GaN-on-Si expected to have more of an impact on price-sensitive markets such as wireless base station amplifiers, satellite communication and cable television(CATVs).[9,10] In view of the potential held out by GaN-on-Si electronics, a thorough understanding and control of all aspects affecting their device performance is timely and of utmost importance. One such issue is the presence of a parasitic channel at the epitaxial nitride-silicon interface. Substrate losses are a major concern for RF applications due to eddy currents induced by capacitive coupling between the substrate and channel at such high frequencies (2.4 GHz in wireless base stations, for instance), which is the reason GaN-on-Si HEMTs for high-frequency operation are manufactured on high resistivity (>10,000 Ω-cm HR-Si) substrates. Hence, the presence of a parasitic conduction pathway at the nitride-silicon interface is detrimental in terms of the switching efficiency and reduces the output power that can be extracted from these devices.[7,11,12] GaN power transistors on silicon, on the other hand, switch at more modest frequencies (~1MHz) which reduces the impact of substrate switching losses and are commonly fabricated on low resistivity Si substrates. However, the presence of a parasitic channel in power devices has been shown to contribute to higher leakage currents and lead to accelerated breakdown, effectively reducing the off-state blocking voltages of these devices.[13-16] Substrate parasitic channel formation is therefore a pressing technological issue affecting the performance and reliable operation of GaN-on-Si transistors for both RF and power device applications. In the following sections, we first examine the possible sources of the parasitic substrate channel and follow it up with our experimental results, Discussions and Conclusions.

## 2. ON THE ORIGIN OF THE PARASITIC SUBSTRATE CHANNEL

The physical origin of the parasitic substrate channel can be traced back to the growth of III-nitride stacks on silicon. The epitaxy of GaN on Si universally begins with an AlN nucleation layer due to the melt-back etching observed for direct growth of GaN on Si.[17-19] This is followed by the introduction of stress-mitigating transition layers for defect and stress management and subsequently, the GaN buffer and the active device layers including the AlGaN/GaN heterojunction which constitute the HEMT.[20] AlGaN/GaN interfaces demonstrate extremely high two-dimensional electron gas (2DEG) densities ($\sim 10^{13}$ cm$^{-2}$) due to the difference in spontaneous and piezoelectric polarization between these two materials,[21] while the source of these electrons is widely attributed to donor-like surface states which populate the potential well.[22] In this context, it is well worth pointing out that nitride growth on silicon inherently gives rise to a dissimilar semiconductor-semiconductor hetero-interface at the AlN/Si junction. AlN is a wide band gap semiconductor and possesses the highest value of spontaneous and piezoelectric polarization among the family of group III-A nitrides and hence leads to a very large "polarization step" at the interface with non-polar Si substrates. Typically, MOCVD growth of AlN on Si gives rise to Al-polar material which leads to a positive fixed charge density at the interface. The magnitude of spontaneous polarization in single crystalline AlN is 0.081 C/m$^2$, corresponding to an interfacial charge density of $5 \times 10^{13}$ cm$^{-2}$ for the case of an ideal defect-free interface.[23] We note that for good quality material, the nature of such an interface is clean and abrupt, devoid of any amorphous interfacial layers.[19,24] Furthermore, AlN films are strained in tension during growth on Si and hence any piezoelectric polarization field is expected to add to the existing spontaneous polarization component.[12,21,25] However, the severe lattice and thermal mismatches between AlN and Si give rise to c-axis oriented AlN domains of 50-100 nm which then coalesce to form a defective AlN film, with a dislocation density of $10^{13}$ cm$^{-2}$ typically observed in these layers.[25,26] These defective regions could reduce the magnitude of polarization charge at the interface by disrupting the epitaxial registry of nitride films. The presence of an interfacial fixed positive charge due to polarization differences will lead to an inversion layer of electrons in the Si substrate even for no applied bias. However, the mobility of these mobile interfacial carriers is expected to be poor due to defects and imperfections at the interface acting as trap states in silicon. An inversion layer of electrons at the AlN/Si interface attributed to the polarization in AlN with surface donors contributing these electrons has been inferred previously in literature, with Yacoub et al. extracting an electron density of $3 \times 10^{13}$ cm$^{-3}$ without freeze-out at temperatures down to 77K,[14] while Luong et al. have measured a higher RF transmission line loss for thicker AlN films on HR-Si as compared to thinner ones, attributed to higher interfacial electron densities.

On the other hand, the growth of III-nitrides also involves the introduction of Al and Ga containing precursors into the reaction chamber at extremely high temperatures (>900°C for MOCVD layers) which could lead to auto-doping of the exposed silicon surface by Al/Ga atoms, which are acceptors in Si and hence dope the substrate p-type. The presence of significant stresses in these films (>1GPa)[26,27] and extended line defects such as dislocations could also serve as localized pathways for the diffusion and segregation of Al and Ga from nitride films to the substrate. Contrary to the inversion layer of electrons discussed before, this mechanism would give rise to mobile holes and immobile ionized acceptors in Si. Such a parasitic channel due to Ga (and to a lesser extent Al) diffusion into the Si with a surface concentration $>10^{17}$ cm$^{-3}$ and their effect on RF transmission line losses has also been reported.[11]

We thus have two plausible but entirely distinct mechanisms for parasitic channel formation, both of which have been experimentally observed, and which can operate either individually or together contributing to observed RF and power performance degradation. Additional factors such as the diffusion of other growth species during the epitaxy of nitride films into silicon (N, C and O to name a few), and contributions from dopant re-distribution in Si substrates themselves during the high growth temperatures, can also affect parasitic channel formation. But the effects of these other factors have not hitherto been observed or explored.

From the above discussion, we see that the exact mechanisms underlying the formation of the substrate parasitic conduction channel are quite involved and need careful and systematic investigation to unravel. As a first step in this direction, this paper focuses solely on the role of AlN nucleation layers in parasitic channel formation for MOCVD grown epitaxial nitride stacks on Si. Using a combination of C-V and conductance spectroscopy we show that the parasitic channel formed at the AlN/Si interface is due to increased acceptor density at the surface of the substrate and it demonstrates a rather striking dependence on the thickness of the AlN films grown. Thicker AlN films display both a higher acceptor density in Si and a higher trap density at the AlN/Si interface. Secondary-ion mass spectrometry (SIMS) results are presented and correlated with electrical measurements, which show that the formation of thermal acceptor complexes in silicon are the proximate cause for parasitic channel formation at these AlN/Si interfaces. Additionally, low-temperature magnetoresistance measurements reveal a transition from positive to negative magneto-resistance with increasing AlN thickness, with a square dependence at low magnetic fields indicative of two-dimensional carrier confinement due to an inversion layer of electrons at the interface.

## 3. RESULTS AND DISCUSSION:

### Capacitance-Voltage Measurements of AlN/Si Interfaces

Capacitance measurements are one of the most versatile methods to characterize semiconductor interfaces and have been extensively applied to nitride thin films.[28,29] Since the use of HR-Si substrates makes CV measurements challenging due to additional substrate capacitance contribution, p-Si substrates (~5 Ω.cm) were used for these experiments. Al/AlN/p-Si capacitors were fabricated with four different AlN film thicknesses - 50, 100, 150 and 200 nm.[30] The measured high frequency C-V curves (1 MHz) for all four AlN thicknesses normalized to their dielectric capacitances are as shown in Figure 1. Also plotted alongside are the ideal high-frequency MOSCAP C-V curves, calculated for p-type Si samples of the same doping density used in the experiments.[28] The three notable features that can be observed from these C-V curves are the following:

- Hysteresis between forward and reverse voltage sweeps.
- Stretch out of the measured C-V as compared to the ideal case.
- The deviations in depletion and inversion capacitance from the ideal case observed for the 150 nm and 200 nm AlN samples.

Both the anti-clockwise hysteresis loops and the stretch-out in the CV indicate the presence of trap states at the interface. An estimate of the interface trap density from the hysteresis can be obtained by estimating the shift in the flat-band voltages for the forward and reverse sweeps i.e $N_{it} = C_{ox}\Delta V_{FB}/q$, where $N_{it}$ is the interface trap density in cm$^{-2}$, $C_{ox}$ the dielectric capacitance at accumulation and $\Delta V_{FB}$ the shift in flat-band voltage.

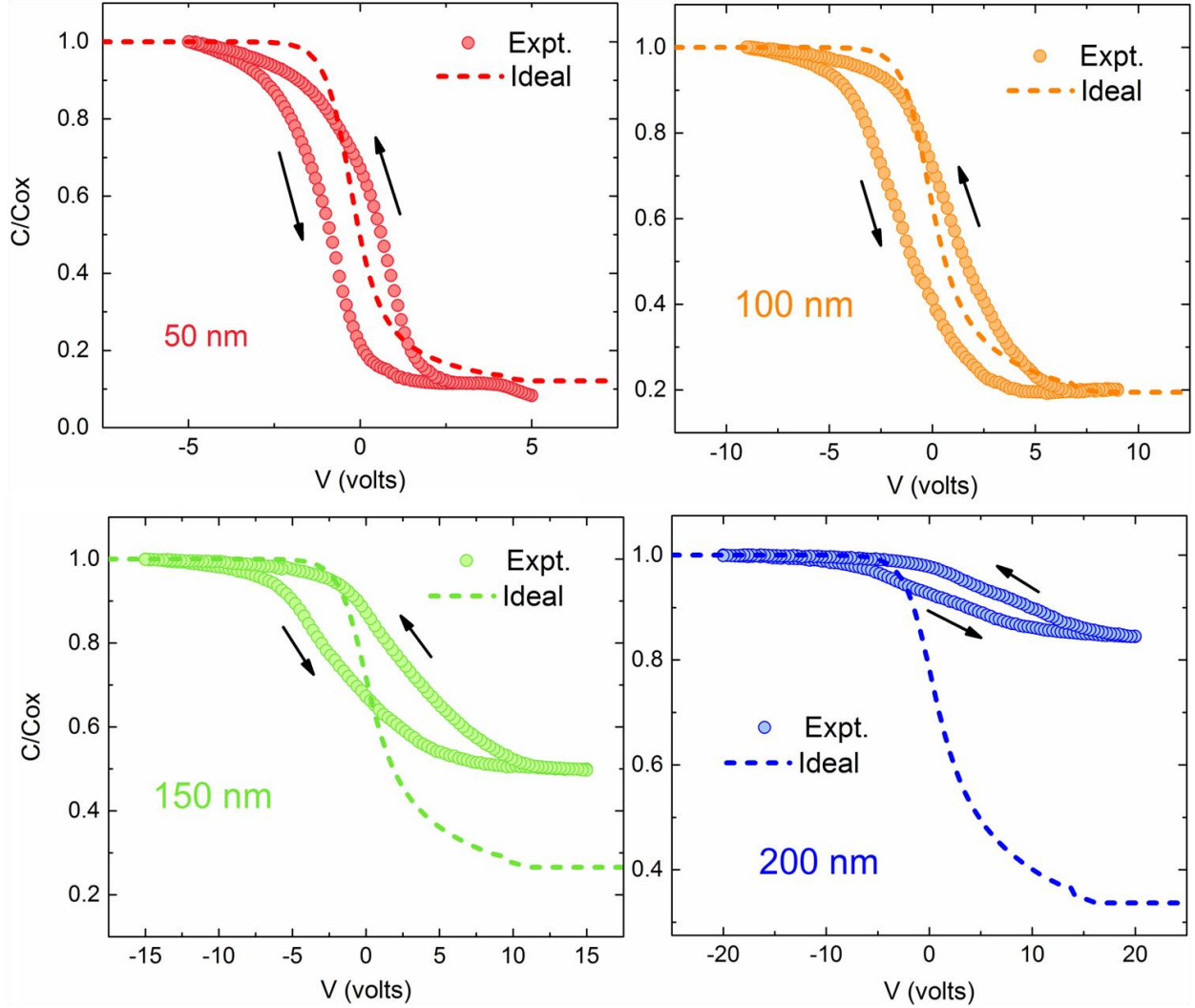

**Figure 1.** Normalized 1 MHz C-V curves of AlN/p-Si capacitors for the four AlN thicknesses - 50, 100, 150 and 200 nm - investigated in this study. Also shown are the ideal high-frequency C-V curves for each thickness for comparison. The 150 nm and 200 nm capacitances show significant deviations from the ideal behavior at positive bias voltages for reasons discussed in the text.

The presence of interface traps can be further quantified by the stretch out in the measured 1 MHz C-V curves when compared to ideal high frequency C-V characteristics. The interface trap density can be extracted from the high-frequency C-V curves by employing Terman's method i.e,

$$D_{it} = \frac{C_{ox}}{q}[(\frac{d\psi_s}{dV})^{-1} - 1] - \frac{C_D}{q} \qquad (1)$$

where $C_{ox}$ stands for the dielectric capacitance, $C_D$ the depletion capacitance, and $\psi_s$ the surface potential.[28] The interface trap densities for the different AlN thicknesses as extracted from the hysteresis and Terman's method are summarized in Table 1. We see that these magnitudes compare favorably with reported values for high-k/Si interfaces. The large deviations of the measured capacitance from the ideal values at inversion for the 150 and 200 nm films prevent the accurate estimation of trap densities.

| AlN film thickness (nm) | Trap Density from Hysteresis $C_{ox}\Delta V_{FB}/q$ (cm$^{-2}$) | Trap Density at midgap from Terman's method (cm$^{-2}$) |
|---|---|---|
| 50 | 1.26x10$^{12}$ | 1.74x10$^{12}$ |
| 100 | 1.22x10$^{12}$ | 2.07x10$^{12}$ |
| 150 | 1.33x10$^{12}$ | - |
| 200 | - | - |

**Table 1.** Interface trap densities, for the four different AlN thicknesses, as extracted from the hysteretic shift in flat band voltage and Terman's method. Trap densities for the thicker samples cannot be extracted as the inversion capacitance deviates significantly from its expected value as discussed in the text, and in the 200 nm case, observed $C_{min}>C_{FB}$.

Lastly, the thicker samples clearly show deviations from the ideal C-V case, in the form of much higher capacitances in the depletion and inversion region. While this effect is absent in case of the 50 and 100 nm samples, it begins to manifest in the 150 nm case. In case of the 200 nm, the capacitance in the inversion region is much higher than even the expected flat band capacitance for this sample. This feature cannot be explained merely by the presence of an inversion layer of electrons at this interface. Two possible reasons for such an increase in capacitance are the presence of a very high density of interfacial traps or increased p-type doping of the Si close to the AlN/Si interface. However, given the estimates for trap densities extracted above and a < 5% dispersion in $C_{max}$ and $C_{min}$ observed over a 1 kHz-1MHz frequency range, it is unlikely that trap states are the cause of this feature. Trap densities extracted from a conductance technique on these samples also reveal that the interfacial density does not increase significantly to justify such as assertion and will be discussed next. This suggests that the increased acceptor concentration at the surface of Si, which would lead to a narrower maximum depletion width and increased $C_{min}$, is the cause of the observed deviation from the expected CV curves for thicker AlN samples. Assuming that the up-shift in $C_{min}$ is due to increased surface p-type doping, the extent of such doping is estimated to be $5.07 \times 10^{17}$ cm$^{-3}$ from the value of the minimum capacitance, while the doping density of the starting Si substrate used was $2.8 \times 10^{15}$ cm$^{-3}$. The interface trap density ($D_{it}$) value for the 200 nm film can now be estimated using Terman's method with this increased doping density and is found to be $3 \times 10^{12}$ cm$^{-2}$ at midgap. This is larger than the trap densities observed for the thinner films, but not significantly high enough to increase $C_{min}$ by itself, and again points to increased surface p-type doping in Si.

**Conductance Spectroscopy Measurements of AlN/Si Interfaces**

To further elucidate the electrical properties of the AlN/Si interface and their thickness dependence, conductance spectroscopy was used to quantify the interfacial trap densities and their characteristic time constants of these samples. The normalized conductance ($G_p/\omega$) is related to the measured capacitance ($C_m$) and conductance ($G_m$) as

$$\frac{G_p}{\omega} = \frac{\omega G_m C_{ox}^2}{G_m^2 + \omega^2 (C_{ox} - C_m)^2} \qquad (2)$$

where $\omega$ and $C_{ox}$ are the angular measurement frequency and capacitance of the gate dielectric respectively. In case of a continuous distribution of interface states in energy throughout the band gap, the normalized conductance is in turn related to the interface trap density $D_{it}$ as

$$\frac{G_p}{\omega} = \frac{qD_{it}}{2\omega\tau_{it}}\ln[1+(\omega\tau_{it})^2] \qquad (3)$$

where $\tau_{it}$ is the interface trap time constant ($R_{it}*C_{it}$). The expression for $D_{it}$ can further be approximated in terms of conductance maxima as $D_{it} = 2.5/q *(G_p/\omega)$.[31] For the sake of brevity, we restrict our discussions to the conductance spectra of the 50 nm and 200 nm samples, which are as shown in Figure 2.

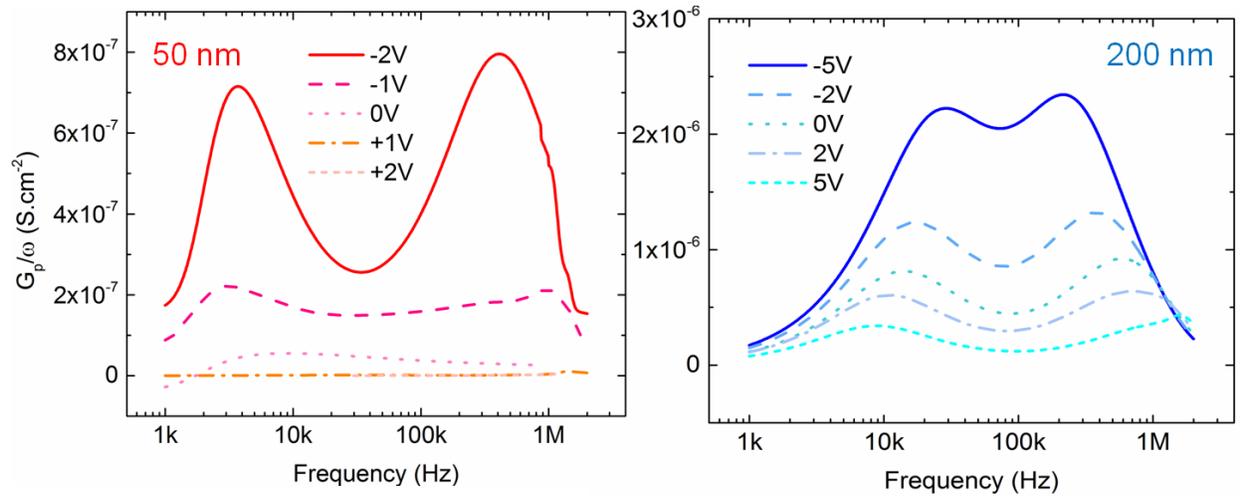

**Figure 2.** Normalized conductance ($G_p/\omega$) as a function of frequency for the 50 nm (left) and 200 nm (right) thick AlN films on p-Si, swept from depletion into weak inversion. Both samples show two distinct traps corresponding to slow and fast trap states with the 200 nm sample having ~3x the trap densities as the 50 nm AlN film.

We see that two distinct slow and fast trap states are observed for both film thicknesses, having similar time constants that vary with applied gate bias, which indicates that increasing the AlN film thickness does not give rise to additional traps. However, the density of existing traps is exacerbated as can be observed from the higher $G_p/\omega$ values for the 200 nm thick films compared to the 50 nm sample. Figure 3 plots the trap densities and characteristic time constants as a function of applied bias for the 50 and 200 nm AlN films. For the 50 nm AlN films, the slow traps (denoted by $\tau_{it,1}$, $D_{it,1}$) have a maximum time constant of $4.23 \times 10^{-5}$ s with a zero bias $D_{it}$ of $8.6 \times 10^{11}$ cm$^{-2}$eV$^{-1}$ while the minimum time constant and zero bias $D_{it}$ for the fast traps ($\tau_{it,2}$, $D_{it,2}$) are $3.77 \times 10^{-7}$ s and $4.1 \times 10^{10}$ cm$^{-2}$eV$^{-1}$. The trap response time of the fast traps shifts outside the measurement frequency range for $V_G > 0$V and hence cannot be further quantified. Similarly, the slow traps in the 200 nm AlN films exhibit a maximum time constant of $6.2 \times 10^{-6}$ s and a $D_{it}$ of $1.27 \times 10^{13}$ cm$^{-2}$eV$^{-1}$ at zero applied bias and fast traps have a minimum time constant of $7.5 \times 10^{-7}$s and a zero bias $D_{it}$ of $1.44 \times 10^{13}$ cm$^{-2}$eV$^{-1}$. We also see that the defect density as estimated from Terman's method earlier provides but a lower limit for $D_{it}$ as only the trap states capable of responding at the C-V measurement frequency of 1 MHz contribute to the stretch-out whereas the conductance method profiles the trap density both as a function of frequency and applied bias. These results represent the first actual characterization of trap states at the AlN/Si interface, which are expected to impact high-frequency device performance and reliability of AlGaN/GaN HEMTs, and indicate that passivation schemes, either in-situ or ex-situ, need to be explored to reduce the trap densities at this interface.

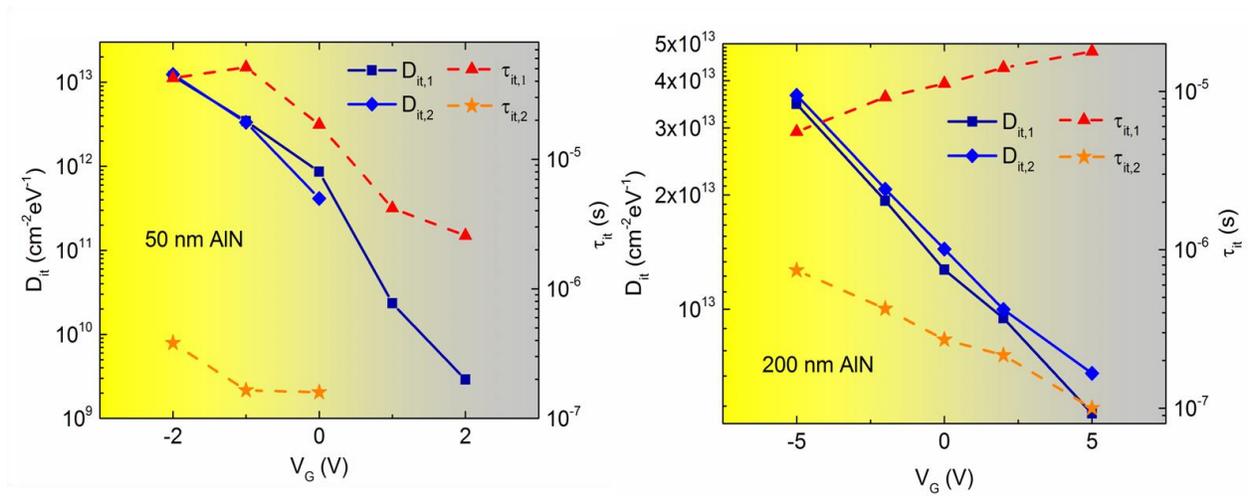

**Figure 3.** Evolution of $D_{it}$ and $\tau_{it}$ in the depletion region with gate voltage for 50 nm (left) and 200 nm (right) AlN/p-Si capacitors. Slow and fast traps are denoted by suffixes 1 and 2 respectively. Higher trap densities are observed in the case of the thicker AlN film while the time constants remain similar.

**The Physical Origin of Increased Surface Acceptor Concentration:**

We now address the origin of increased acceptor densities at the surface of the Si substrate for thicker AlN films. Two possible reasons which have already been mentioned are auto-doping of Si substrate with Al-containing species either during growth or from the AlN film and dopant pile-up at AlN/Si interfaces from the acceptors in the p-Si substrate itself (boron, in this case). In case of GaN growth on AlN nucleation layers, the diffusion of Ga into Si is another factor that needs to be accounted for. Additionally, the presence of nitrogen, carbon, and oxygen (in trace amounts from reactants and in the Czochralski-grown Si substrates) which are invariably present during MOCVD growth of nitride films also get incorporated in Si when subject to the high growth temperatures. The AlN films thus grown are significantly strained in tension with growth stresses >1GPa[19,25] and an increase in AlN thickness hence increases the strain energy built up in these films. This increased strain energy could potentially create new defects at the interface which in turn can act as sinks for boron pile-up from the bulk silicon substrate, as has been observed in literature for ion-implantation induced damage in p-Si.[33-35] Also, the dislocation density in AlN nucleation layers is ~$10^{13}$ cm$^{-2}$, which can act as diffusion pathways for the migration of Al or N species from the AlN film to the substrate.[36] In order to distinguish between these possible explanations of the increased surface acceptor densities, secondary ion mass spectroscopy (SIMS) was performed on the AlN/p-Si films. Figure 4 shows the depth profile of the concentration of boron atoms as measured from the top of the AlN/p-Si samples. We see that the concentration of boron, originating from the uniformly doped substrate, peaks at the AlN/Si interface and the magnitude of this peak reduces with increasing AlN thickness. This occurs due to boron diffusion into thicker AlN films due to their exposure to higher temperatures for longer durations needed to grow thicker films. Peak boron concentrations observed can reach almost $10^{19}$ cm$^{-3}$ as compared to the starting substrate doping density of <$3\times10^{16}$ cm$^{-3}$. The thinner AlN films have a slightly higher boron concentration near the Si surface, both in terms of peak boron concentrations and the integrated areal density of boron in Si. While thicker films are exposed to the higher growth temperature for longer times (300 seconds of additional exposure duration at 1050°C for the 200 nm AlN sample compared to the 50 nm sample), prior reports on the activation of ion-implanted boron in crystalline Si indicates that such a time difference is insufficient to significantly impact the activation of boron atoms at this temperature.[37-40] Hence the presence of higher surface boron concentrations cannot be the reason for the higher surface acceptor concentration observed for thicker films. Furthermore, the fact that segregation of boron to the AlN/Si interface for thinner samples does not manifest in their capacitance profiles indicates that they are electrically inactive due to their non-substitutional nature and the formation of clusters/complexes.[41]

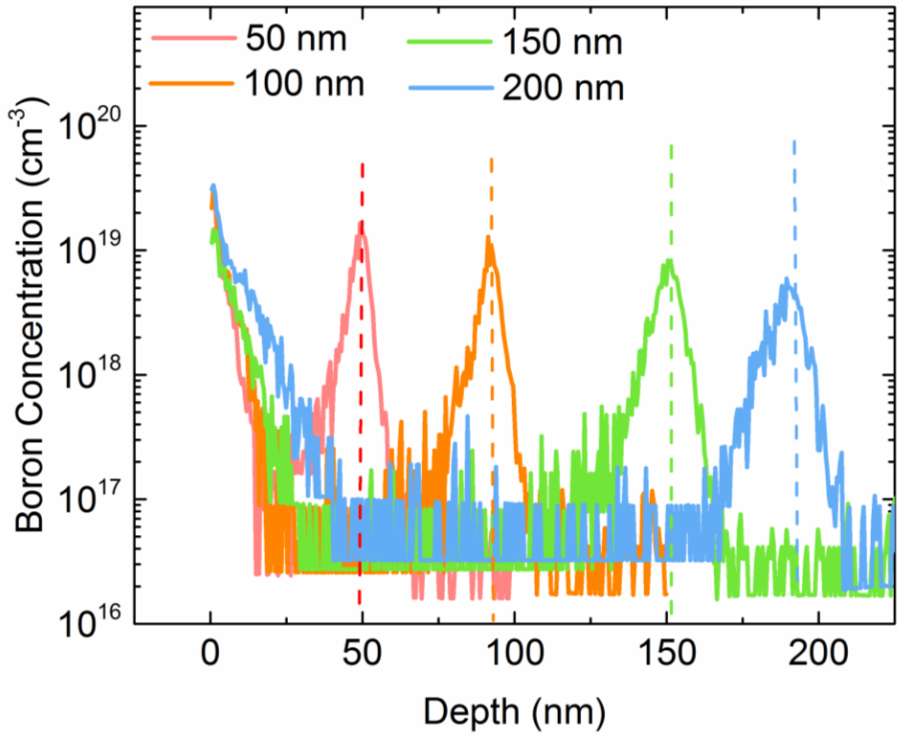

**Figure 4.** SIMS profile of boron concentration as a function of depth for the four different AlN film thicknesses. The dotted lines indicate AlN/p-Si interfaces. The peak concentrations and the integrated areal density of boron atoms in Si are higher for thinner films as diffusion into AlN itself increases with film thickness.

Al diffusion into Si during the growth of AlN could be another reason for the increased acceptor concentration. Figure 5 show the SIMS profile for Al concentration for the 50 and 200 nm AlN films measured through the backside of the sample after thinning the silicon substrate. The observed concentration of Al is also >$10^{18}$ cm$^{-3}$ and a higher tail concentration of Al can be seen into Si. However, no significant difference in the Al concentration between 50 and 200 nm films are observed in terms of peak and areal Al densities in Si which makes it unlikely that auto-doping of Si with Al-containing species is the reason for the observed capacitance effects, analogous to the case of the boron discussed above.

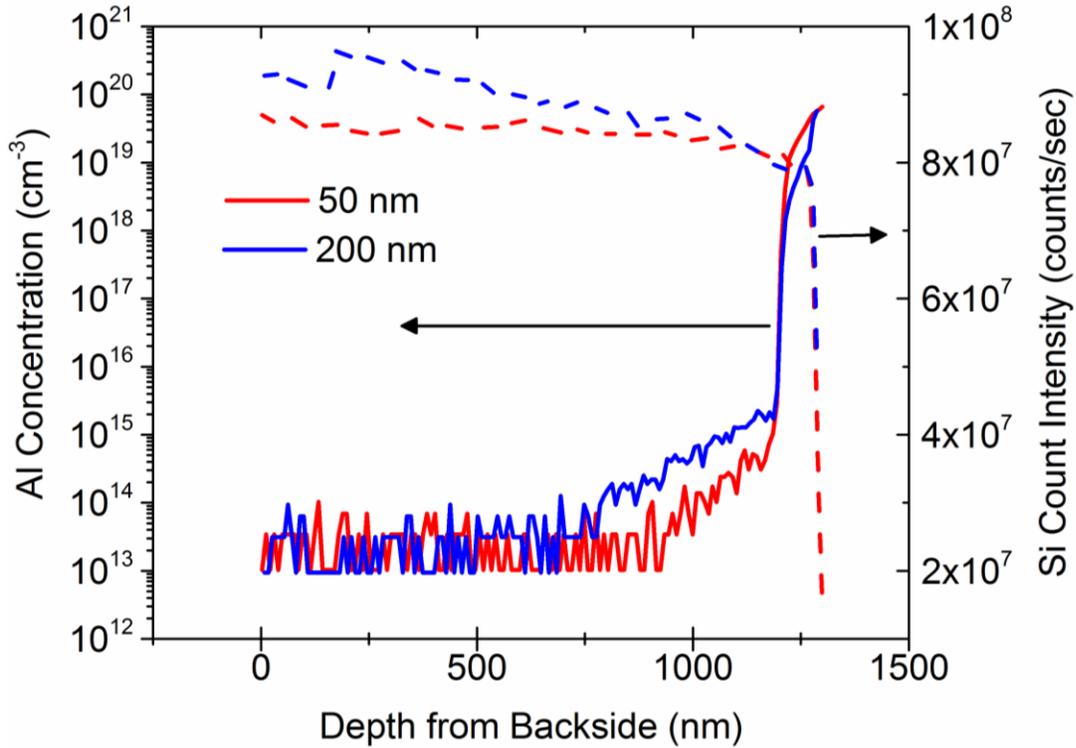

**Figure 5.** SIMS profile of aluminum concentration as a function of depth from the backside of 50 and 200 nm thick AlN/p-Si samples. The dotted lines indicate Si SIMS count intensity which shows the transition from Si to AlN. The peak concentrations and the integrated areal density of aluminum atoms in Si are slightly higher for the 50 nm films.

Given the similar concentrations of boron and Al in the thinner and thicker films, it is clear that the source of increased acceptor doping lies elsewhere. Figure 6 shows the SIMS profile of the other two pre-dominant atomic species during the growth of nitrides, viz; nitrogen and oxygen. Indeed, the observed peak nitrogen concentrations in Si for all the AlN films is at least an order of magnitude higher than any other dopant species, followed by the oxygen concentration in silicon.

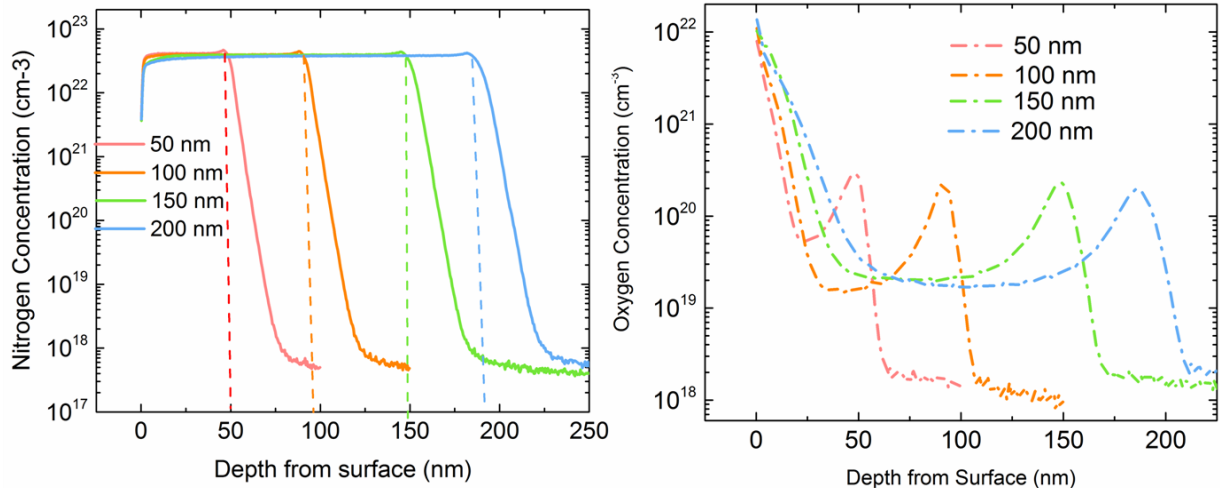

**Figure 6.** SIMS profile of nitrogen (left) and oxygen (right) concentration as a function of depth for four different AlN film thicknesses with peak concentrations occurring at the AlN/p-Si interface. Both the peak concentrations and integrated areal densities are comparable for all four film thicknesses.

Once again, we see that the peak concentrations and integrated areal densities of these species are comparable to each other for the four AlN thicknesses considered here. The behaviour of nitrogen complexes in silicon is quite involved with many nitrogen-interstitial and vacancy complexes predicted to be deep donors and acceptors.[42] Experimentally, however, nitrogen is known to transition from being a donor in Si with <1% activation at low annealing temperatures to being electrically inactive for annealing temperatures >850°C,[43] which should be the case for MOCVD grown AlN films (1050°C in our case, for instance). However, the simultaneous presence of high concentrations of oxygen and nitrogen atoms in the top surface of silicon indicates the possibility of forming Si-O-N complexes which can act as thermal acceptors.[44] Thermal acceptor concentration has been reported in implanted Si samples to first increase with annealing times until it reaches a steady state concentration after a few hours of annealing at high temperatures. Even though the actual oxygen and nitrogen concentrations are not too different in the case of thinner and thicker AlN films, the difference in growth times (300s in case of 200 nm vs 50 nm film) means that the thicker samples are exposed to higher temperatures for longer durations, thus increasing the concentration of thermal acceptors complex formation and correspond to the time scales reported by Yang et al. for increased thermal acceptor formation.[44] We note that even a 0.02% activation of N-containing complexes thus formed is enough to account for the increased acceptor concentrations observed in the capacitance profiles. Such thermal acceptor formation is expected to be exacerbated during the growth of a complete GaN HEMT stack due to the higher growth times (typical growth rates ~1µm/hr), while the effects of Ga-diffusion from the stack into silicon would also need to be accounted for.

**Low-temperature Magnetoresistance Behavior with AlN Film Thickness:**

The preceding discussion makes it clear that the pre-dominant thickness dependent parasitic channel effects observed at AlN/Si interfaces are due to thermal acceptor complexes formed in Si during film growth. However, the presence of an inversion layer of electrons close to the interface, which is swamped out by the effect of the increased acceptor concentration, is still a possibility. *Such an inversion layer, if it exists, is expected to persist down to low temperatures while the acceptor contribution to the parasitic channel are expected to be frozen out.* In order to investigate this possibility, samples were fabricated in van der Pauw Hall geometry using Indium contacts and were subjected to low temperature magnetoresistance measurements.

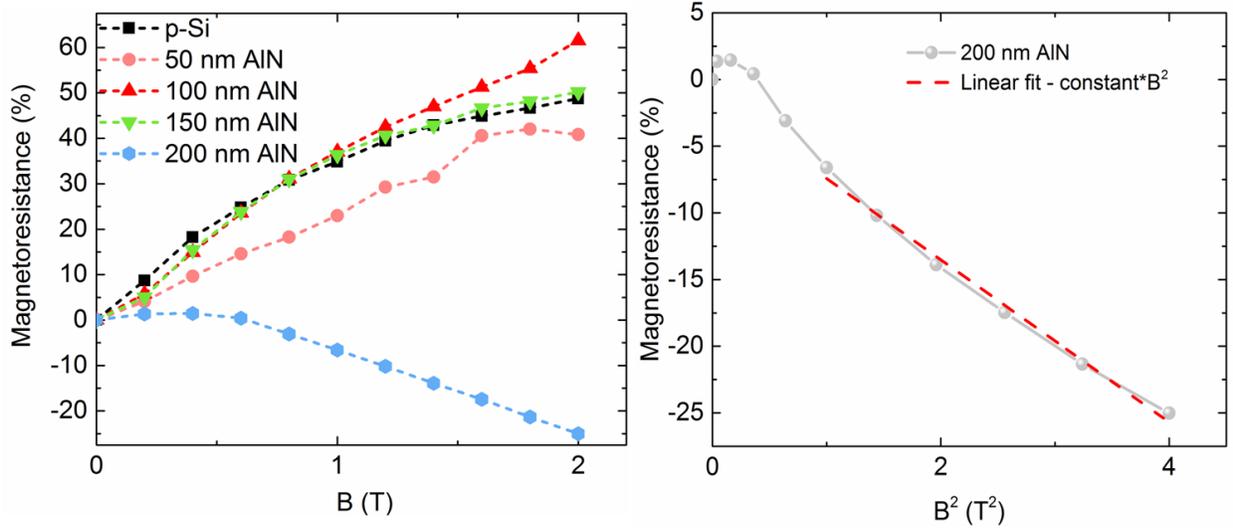

**Figure 7.** (left) Magnetoresistance of four different AlN film thicknesses on Si at 5K as a function of applied magnetic field (B). Also shown for comparison is the MR of a bare p-Si sample having the same doping density as the substrates. (right) Negative MR of the 200 nm AlN sample as a function of square of applied magnetic field ($B^2$). Dotted line indicates a linear fit to $B^2$ for comparison.

The magnetoresistance (MR = $\Delta R_{xx}/R_{xx}$ in %) as a function of magnetic field (B) at 5K for a reference p-Si sample of the same doping density and the four AlN samples under investigation are as shown in Figure 7. We observe that the standard p-Si with no AlN and p-Si deposited with AlN films of thicknesses 50, 100 and 150 nm exhibit large positive MR with applied magnetic fields, which has been attributed in literature to space-charge injection of carriers through the contacts.[45] In contrast, the 200 nm sample shows a high negative magnetoresistance (NMR) of 25% at 2T. The magnetic field dependence due to weak localization in a 2D-confined carrier system is proportional to $B^2$ at low magnetic fields,[46] and such a dependence is also observed for the 200 nm samples as shown in Figure 7. The transition from positive to negative NMR with increasing AlN thickness is further indicative of a transition from 3D to 2D carrier confinement for these samples and such an observation has also been reported for in-plane magneto-resistance changes in LAO/STO hetero-junctions, where the transition from a 3D to a 2D electron gas engenders a change from positive to negative MR due to the stronger localization effects with reduced dimensionality.[47] These features are suggestive of two-dimensional confinement at the AlN/Si interface in case of the 200 nm samples and they support the possibility of the formation of an inversion channel of electrons at the AlN/Si interface for the thicker samples, due to the polarization discontinuity at AlN/Si and contribution from surface donors as reported by Luong et al.[12] Further investigation of magneto-transport at AlN/Si interfaces is necessary to conclusively demonstrate such confinement effects and offers an interesting topic for future studies.

In conclusion, thickness-dependent parasitic channel formation at AlN/Si interfaces is shown to be composed of both electron and hole contributions, from the polarization difference at AlN/Si and surface donors, and increased thermal acceptor formation due to Si-O-N complexes at Si surface respectively. At room temperatures, the effect of the surface acceptors completely dominates the characteristics of the interface and manifests in the thickness dependent change in capacitance of these samples. SIMS analysis reveals that thermal acceptor formation due to Si-O-N complexes is responsible for the increased surface acceptor densities for the thicker films, which also demonstrate higher interfacial trap densities as compared to thinner films. Low temperature magnetoresistance measurements at 5K, well below the acceptor freeze-out temperature reveal a transition from positive to negative MR for thicker samples, with a $B^2$ dependence for NMR, indicative of two-dimensional confinement due to an inversion layer of electrons at the AlN/Si interface. The use of thinner AlN layers is expected to help suppress both effects, but its effect on the properties of the rest of the layers constituting the HEMT stack needs to be considered as well.

## METHODS:

### MOCVD growth of AlN and device fabrication:

AlN films were grown on p-Si substrates (1-10 Ω.cm, 2" prime wafers) using an AIX 200/4 RF-S MOCVD system with trimethyl aluminum and ammonia as precursors, after a HF pre-treatment,[19] with a growth rate of 25 nm/min and a V/III ratio of 434. 50, 100, 150 and 200 nm AlN films were grown and film thickness was monitored in-situ at the centre of the wafer to ensure accurate thicknesses. Ellipsometry of these four films post-growth indicated average film thickness to be 54 nm, 97 nm, 159 nm and 202 nm respectively with a 5% variation in thickness across the 2" wafer.[30] AlN/p-Si MOSCAP fabrication involved deposition of circular Al electrodes of 300 μm diameter using thermal evaporation of the top-side with a blanket Al metallization of the bottom forming the back electrode. The samples were then annealed in forming gas at 400°C, 15 min for post-metallization annealing. For the magnetoresistance measurements, van der Pauw squares of ~1 cm were used with Indium contacts soldered onto the four corners at 400°C.

### Electrical and SIMS characterization:

AlN/Si MOSCAPs were then characterized using an Agilent 4941 Impedance Analyzer for capacitance and conductance measurements over a frequency range of 1 kHz to 2 MHz with an a.c. excitation of 30 mV. A frequency dispersion of ~5% was observed from 1 kHz to 1 MHz with the dissipation factors less than 0.1. Magnetoresistance measurements were performed using a Lakeshore CRX-VF closed cycle He-cryostat at 5K with a superconducting magnet of maximum field 2T perpendicular to the sample plane. SIMS measurements, both front-side and back-side with substrate polished down to ~1μm, were performed at Evans Analytical Group Inc., USA.

**ACKNOWLEDGEMENTS**

We acknowledge the Ministry of Defence, Government of India through sanction number TD-2008/SPL-147 and the Minsitry of Electronics and IT Government of India thorugh Centre of Excellence in Nanoelectronics project for funding to carry out this research, the National Nano Fabrication Centre (NNFC) and the Micro and Nano Characterization Facility (MNCF) of the Centre for Nano Science and Engineering, Indian Institute of Science for providing access to the fabrication and characterization facilities. H.C. wishes to thank Prof. V Venkataraman, Department of Physics, IISc, Bangalore for helpful discussions.

**Author contributions statement**

N.B., H.C. and S.R. conceived the experiments. H.C. performed the MOCVD growth, device fabrication and characterization, and wrote the paper. N.B. and K.N.B supervised the electrical characterization and analysis, and R.M. the magnetoresistance measurements. All authors discussed and analyzed the results and interpretation and reviewed the paper.

**Additional Information**

**Competing financial interests**

The authors declare no competing financial interests.